\documentclass[12pt,a4paper,twoside]{article}
\usepackage{graphics,epsfig}
\newcommand{\dd}{\mathrm{d}}
\newcommand{\sigT}{\sigma_{\mathrm{T}}}
\newcommand{\sigL}{\sigma_{\mathrm{L}}}
\newcommand{\sigh}{\sigma^{\mathrm{h}}}

\newcommand{\sigtot}{\sigma_{\mathrm{tot}}}
\newcommand{\FT}{F_{\mathrm{T}}}
\newcommand{\FL}{F_{\mathrm{L}}}
\newcommand{\FLh}{F_{\mathrm{L}}^{\mathrm{h}}}
\newcommand{\FLp}{F_{\mathrm{L}}^{\mathrm{p}}}
\newcommand{\alphas}{\alpha_{\mathrm{s}}}
\newcommand{\Ecm}{E_{\mathrm{cm}}}
\newcommand{\pT}{p_{\perp}}
\newcommand{\pL}{p_{\|}}
\newcommand{\Tbar}{\overline{T}}
\newcommand{\ymin}{y_{\mathrm{min}}}
\newcommand{\gtrsim}{\raisebox{-0.8mm}%
{\hspace{1mm}$\stackrel{>}{\sim}$\hspace{1mm}}}
\newcommand{\lessim}{\raisebox{-0.8mm}%
{\hspace{1mm}$\stackrel{<}{\sim}$\hspace{1mm}}}
\title{
\begin{flushright}
\normalsize 
LU TP 01-13\\
LUNFD6/(NFFL-7194)2001\\
\vspace*{2cm}
\end{flushright}
Hadronization corrections to helicity components of the fragmentation function}
\author{T.~Sj\"{o}strand$^1$, O.~Smirnova$^2$\thanks{\emph{On
      leave from:} JINR, 141980 Dubna, Russia}~~and Ch.~Zacharatou
  Jarlskog$^2$}

\date{\small $^1$Department of Theoretical Physics, Lund University,
    S\"{o}lvegatan 14A, S~22362 Lund, Sweden \\ $^2$ Department of
    Elementary Particle Physics, Lund University, P.O.Box~118, S~22100
    Lund, Sweden}
\begin{document}
\hyphenation{had-ron had-rons had-ro-ni-za-tion had-ro-ni-zate had-ro-nize}

\maketitle
\abstract{In the hadronic decays of $Z^0$, gluon emission leads
  to the appearance of the longitudinal component of the fragmentation
  function, $F_L$. Measurement of $F_L$ and the transverse component,
  $F_T$, could thus provide an insight into the gluon fragmentation
  function. However, hadronization corrections at low $x$ can be
  significant. Here we present a method of accounting for such
  corrections, using the \textsc{Jetset} event generator as illustration.
}

\section{Introduction}
\label{intro}
Studies of fragmentation functions have always been important, since
these distributions can not be predicted theoretically, but only be
measured experimentally, and consecutively be described by
phenomenological models. Hadronic decays of $\gamma^*/Z^0$ provide a
particularly convenient set of events for analysis and interpretation.
Helicity components of the fragmentation function, measured in such
events, can be used in various QCD studies, e.g., the extraction of
the gluon fragmentation function, and the evaluation of $\alpha_s$.
However, existing theoretical calculation being restricted to the
perturbative region, hadronization corrections must be taken into
account. In what follows, methods for applying such corrections using
the \textsc{Jetset} event generator~\cite{jetset} as an example, will
be discussed.

Consider the angular distribution in the process
$e^+ e^- \to \gamma^*/Z^0 \to q \overline{q}$ in its rest frame.
Assuming that the final quark and antiquark are not charge-tagged,
i.e. that the forward--backward asymmetry is not accessed,
the cross section can be written as~\cite{HK}
\begin{equation}
\frac{\dd\sigma}{\dd(\cos\theta)} =
\frac{3}{8} \, (1 + \cos^2 \theta) \, \sigT +
\frac{3}{4} \, \sin^2\theta \, \sigL ~.
\label{sigmaparton}
\end{equation}
Here $\sigT$ ($\sigL$) is the cross section associated with a
transverse (longitudinal) gauge boson polarization state with respect
to the $q \overline{q}$ axis, and $\theta$ is the polar angle of a
particle with respect to the incoming lepton axis. To lowest order,
only mass effects contribute to a non-vanishing $\sigL$, but only for
the vector part of the cross section, and there only with a
coefficient $\sigL/\sigT = 2 m_q^2/\Ecm^2$. Even for the $b$ quarks
this gives a negligible $\sigL$ contribution at the energies around
the $Z^0$ peak.  Therefore, $\sigL$ effectively starts in
$\mathcal{O}(\alphas)$ of perturbation theory, associated with the
emission of gluons.

Since partons are not directly observable, one may define a
hadron-level analogue of Eq.(\ref{sigmaparton})~\cite{altarelli},
\begin{equation}
\frac{\dd^2\sigh}{\dd x \, \dd(\cos\theta)} =
\frac{3}{8} \, (1 + \cos^2 \theta) \, \frac{\dd\sigT}{\dd x} +
\frac{3}{4} \, \sin^2\theta \, \frac{\dd\sigL}{\dd x} ~.
\label{sigmahadron}
\end{equation}
Here $x$ would preferably be associated with the energy fraction
taken by a hadron, $x_E = 2 E/\Ecm$, so that $\sum x_E = 2$ in
each event. Experimentally it is more convenient
to use the momentum fraction $x_p$. The
transverse and longitudinal fragmentation functions are defined
by a normalization to the total cross section
$\sigtot = \sigT + \sigL$~\cite{NW},
\begin{equation}
\FT(x) = \frac{1}{\sigtot} \frac{\dd\sigT}{\dd x} ~,~~~~
\FL(x) = \frac{1}{\sigtot} \frac{\dd\sigL}{\dd x} ~.
\label{FTL}
\end{equation}

The former is dominated by the fragmentation of quark jets,
whereas the latter receives a major contribution from gluon
fragmentation. Therefore an experimental determination of
$\FL(x)$ is a first step towards an extraction of the gluon
fragmentation function, alternative to what is offered by more
direct methods in three-jet events~\cite{3j-exp}. Several
experimental $\FL(x)$ studies have also been presented~\cite{fl-exp}.

\begin{figure}[ht]
\centering
\mbox{\epsfig{file=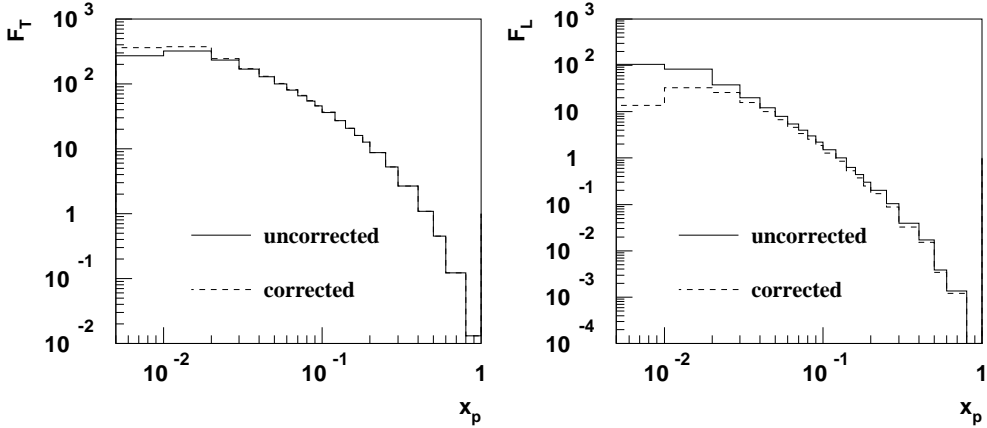,width=0.9\linewidth}} \caption{\it
$F_T(x_p)$ and $F_L(x_p)$ for corrected (smeared) hadron angles (dashed lines) and for
uncorrected (non-smeared) hadron angles (solid lines).}
\label{fig:smear}
\end{figure}

A complication is that hadrons are not moving in the direction
of their imagined mother parton. Already in lowest order of
perturbation theory, for $q \overline{q}$ two-jet events,
nonperturbative hadronization gives an effective $\pT$ smearing
that induces a nonvanishing $\FL(x)$ even where none is expected.
Furthermore, the association of a hadron to a single mother
parton is not in agreement with our current best understanding
of the hadronization process, where it is rather the colour field
between a colour-connected pair of partons (a string piece~\cite{string}, or a cluster~\cite{cluster}) that mediates the
hadron production. Therefore the structure of smearing effects
may become rather nontrivial. Obviously, the effects are
especially important at small $x$, which is also the region where
one would hope to have some sensitivity to the gluon fragmentation
function. The string picture also casts in doubt the concept of
a gluon fragmentation function defined from inclusive quantities,
since the string fragmentation of a parton depends on the angles
to other colour-connected partons.

The issue of hadronization corrections to fragmentation functions
was addressed in~\cite{NW,power}. The emphasis was on the $\sigL$ that
can be extracted from $\int_0^1 \FL(x) \, x \, \dd x$ rather than
on $\FL(x)$ itself, however. Therefore we here address how
hadronization affects $\FL(x)$ (and $\FT(x)$). One main conclusion
is that a simple smearing approach is not sufficient to describe
hadronization effects. Thus it appears impossible to define
a completely model-independent,
had\-ro\-ni\-za\-tion-smearing-corrected $\FL(x)$, that could be used
to extract a gluon fragmentation
function. We further suggest a correction procedure, based on
a cluster search strategy, that should give a less model dependent
$\FL(x)$, but at the price of introducing the cluster resolution
scale $y$ as a new parameter in the problem.

\section{The simple smearing}

The string model description of $q \overline{q}$ events introduces
a Gaussian transverse momentum smearing of primary hadrons,
$\propto \exp(-\pT^2/2 \sigma^2) \, \dd^2 \pT$, where
$\sigma \approx 0.36$~GeV~\cite{jetset,LEP2}. Many primary hadrons are
unstable and decay further; this distorts the original Gaussian
spectrum and reduces the average $\pT$.  Since decay products have
smaller $\pL$, momentum parallel to the jet axis, some correlation
is also introduced between $\pT$ and $\pL$. Therefore no simple
parameterization is proposed, but instead a Monte Carlo simulation with
\textsc{Jetset}~7.4~\cite{jetset} is used to histogram the amount
of angular smearing for different $x_p$ bins.

It is now assumed that this smearing should be applied both to quark
and gluon jets, so that a hadron will not move exactly in the direction
of its mother parton. There are obvious shortcomings to
equating different kinds of jets, like that gluon jets have a lower
energy and do not contain decays of charm and bottom hadrons, but those
particular issues only introduce moderate corrections.
More severe objections can be raised to the association of hadrons to
individual partons, as we will discuss further in the next section, but
forget for the moment.

Had particles not been smeared in $\pT$, but parallel with their parton
of origin, the shape of the angular distribution
\begin{equation}
F(x_p, \cos\theta) =
\frac{3}{8} \, (1 + \cos^2 \theta) \, \FT(x_p) +
\frac{3}{4} \, \sin^2\theta \, \FL(x_p)
\label{Fxp}
\end{equation}
in a bin of $x_p$ could be used to extract $\FT(x_p)$ and $\FL(x_p)$
in that bin. The abovementioned smearing will now modify this. The two
angular shapes, $(3/8) \, (1 + \cos^2 \theta)$ and $(3/4) \, \sin^2\theta$,
both normalized to unity, are therefore convoluted with the $x_p$-dependent
smearing distributions, characterized by a distribution in the smearing
angle $\theta^{\mathrm{sm}}$ and an isotropic azimuthal distribution
$\varphi^{\mathrm{sm}}$. That is, a parton at an angle $\theta^{\mathrm{p}}$
will produce a hadron at an angle $\theta^{\mathrm{h}}$, where
\begin{equation}
\cos \theta^{\mathrm{h}} = \cos \theta^{\mathrm{p}} \,
\cos \theta^{\mathrm{sm}} - \sin \theta^{\mathrm{p}} \,
\sin \theta^{\mathrm{sm}} \, \cos \varphi^{\mathrm{sm}} ~.
\label{addangles}
\end{equation}

Data can now be fitted both to the ``non-smeared'' angular
distribution form, Eq.(\ref{Fxp}), and to the convoluted (``smeared'')
analogue.
The resulting distributions for $F_T$ and $F_L$,
obtained using the \textsc{Jetset} generated events,
are shown in Fig.~\ref{fig:smear}.  The effect on $F_T$ and $F_L$ is
visible for hadron momenta below 10\% of the beam energy. The low
momentum region is affected the most, giving $F_L$ values reduced up
to one order of magnitude.

\section{Objections to the simple smearing}

The above smearing procedure is correct to lowest order in $\alphas$,
i.e. it describes how two-jet events can induce a nonvanishing $\FL(x_p)$.
We know, however, that hadronization of three-jet events cannot be described
in terms of a simple incoherent sum of three $q$, $\overline{q}$ and $g$
jets. One example is the string/drag effect~\cite{stringeff,drageff}, i.e.
that particle production is suppressed in the angular region between the
$q$ and $\overline{q}$ and enhanced in the other two regions, well confirmed
experimentally~\cite{exp-drag}. High-momentum hadrons still essentially follow the
separate parton directions, but low-momentum ones are significantly affected.
These are the ones where the angular smearing effects are large to begin with.
It is well-known that the string effect leads to more two-jetlike events, e.g.
in terms of thrust $T$, than implied by symmetric smearing~\cite{stringalphas}.
The reason is to be found in the enhanced production of particles between
two colour-connected partons that are close in angle, leading to them seemingly
being even closer, e.g. that the opening angle between the reconstructed jets
typically is smaller than that between the original partons.
\begin{figure}[ht]
\vspace*{-1cm}
\centering 
\mbox{\epsfig{file=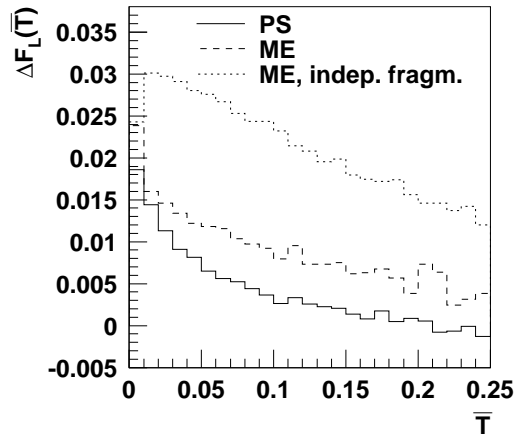,width=0.5\linewidth}}
  \caption{\it Event-by-event hadronization corrections $\Delta
    \FL(\Tbar) = \FLh(\Tbar) - \FLp(\Tbar)$ for different simulations:
    solid line corresponds to \textsc{Jetset}~7.4~PS, dashed --
    \textsc{Jetset}~7.4~ME (both using the string fragmentation), and
    dotted -- \textsc{Jetset}~7.4~ME with independent fragmentation
    scheme.}
\label{fig:toysim}
\end{figure}

There is a nontrivial topology dependence on string effects, especially when
multiple gluon emission is considered. The issue is therefore best studied in
an event generator, by comparing angular distributions on the parton ($i=$p
below) and on the hadron ($i=$h) level. As a simple measure of the jettiness of
events we use $\Tbar = 1-T$, defined on the parton level of each event. The
integrated $x_E$ spectrum is decomposed as
\begin{eqnarray}
\left( \frac{\dd\sigtot}{\dd\Tbar} \right)^{-1} \int_0^1
\frac{\dd^3\sigma^i}{\dd\Tbar \, \dd x_E \, \dd(\cos\theta)}
\, x_E \, \dd x_E &=&~~~~\ \nonumber\\
=\frac{3}{8} \, (1 + \cos^2 \theta) \, \FT^i(\Tbar) +
\frac{3}{4} \, \sin^2\theta \, \FL^i(\Tbar) &&\ ,
\label{sigmaTbar}
\end{eqnarray}
where the $x_E$-weigh\-ting ensures a common normalization
$\FT^i(\Tbar) + \FL^i(\Tbar) = 2$ at parton and hadron level (values
of $\FT^i$ and $\FL^i$ most conveniently are obtained by weighting
each particle with an appropriate angular factor~\cite{NW}). Then
$\Delta \FL(\Tbar) = \FLh(\Tbar) - \FLp(\Tbar)$ is a simple measure of
the hadronization impact on $\FL$. This quantity is shown in
Fig.~\ref{fig:toysim}, for one realistic simulation and two toy ones,
for $u\overline{u}$ events at 91.2~GeV. In the realistic case, a
parton shower is used to generate multiparton configurations, followed
by string fragmentation. The shower develops down to a cut-off scale
$Q_0 \approx 1$~GeV, so that also events in the first bin, $\Tbar <
0.01$, can contain some gluons. The other two histograms are based on
$\mathcal{O}(\alphas)$ matrix elements, where only 2- and 3-parton
configurations are generated, with a cut $\Tbar > 0.01$ on the latter
to avoid the singularities of the 3-parton matrix element. Thus the
first bin here represents pure 2-parton events. While one simulation
is again based on string fragmentation, the other assumes isotropic
smearing around the jet axes, basically the independent fragmentation
scheme of Hoyer~et~al.~\cite{hoyer,stringalphas}. (The same
fragmentation parameters, tuned to the shower model, have been used in
all three cases. A retuning of parameters for the
$\mathcal{O}(\alphas)$ simulations would have given a larger
nonperturbative $\pT$ width $\sigma$ to cover for the lack of
perturbative gluons, and so would have implied even larger
fragmentation smearing.)

The isotropic smearing is, as expected, giving a rather constant
hadronization correction $\Delta \FL(\Tbar)$. There is some jump up
in going from two to three jets that are smeared, followed by a slow but
steady drop with $\Tbar$, since the longitudinal component itself is
increasing in importance with $\Tbar$ and therefore gives an increasing
hadronization smearing of the longitudinal component onto the transverse
one rather than only the other way around. By contrast, the string
fragmentation provides a much steeper drop of $\Delta \FL$ with $\Tbar$,
kicking in immediately when going from two to three partons, and enhanced
in the shower simulation relative to the simpler $\mathcal{O}(\alphas)$
one. At large $\Tbar$ the overall hadronization correction can even turn
negative. Averaging over the $\Tbar$ spectrum (with mean value
$\langle \Tbar \rangle \approx 0.05$), we conclude that the typical
hadronization smearing contribution is only about a third of the naively
expected one, as obtained from two-parton results. (Qualitatively this
agrees with and probably explains a similar observation in~\cite{NW} of
smaller-than-expected hadronization corrections when using
\textsc{herwig}~\cite{herwig}.) That is, if hadronization corrections
are viewed as a power series in $\alphas$, the $\mathcal{O}(\alphas)$
term is of opposite sign and almost as big as the $\mathcal{O}(1)$
one.

It should be remembered, however, that this is integrated over all $x_E$,
and that we have no similar way of addressing results in specific $x$ bins,
since the parton and hadron $x$ spectra are quite different.
Thus it is likely that the $\FL(x_p)$ derived in the previous
section is an underestimation, just like an $\FL(x_p)$ found without any
smearing corrections is likely to be an overestimation, but it appears
impossible to find the ``correct'' $\FL(x_p)$ without making detailed
assumptions about the hadronization process.

\section{Clustering}

Given the problems with the above smearing recipe, we introduce a new
strategy, based on the clustering approach. In a nutshell, we propose
to rotate all hadrons to the direction of the cluster they belong to,
as an approximate way of removing hadronization smearing effects. Only
thereafter is $\FL(x)$ extracted from this modified $\cos\theta$
distribution. The strategy is explained further in the following.

In clustering algorithms, nearby hadrons are combined to form clusters/jets,
in a way that should reflect the underlying partonic state, to some
approximation. The combination process is controlled by (at least) one
separation parameter, call it $\ymin$, such that the final state contains
no pair of clusters closer to each other than that. Clustering algorithms
can be applied also to a partonic state, and here $\ymin$ provides a
regularization of soft and collinear divergences in the perturbative cross
sections. It is then meaningful to calculate the distribution of partons
at a factorization scale $\mu^2 = \ymin \Ecm^2$, and define scale-dependent
fragmentation functions parameterizing the subsequent soft-perturbative and
nonperturbative hadronization. The latter should obey standard QCD
evolution equations, starting from some unknown nonperturbative form at a
low reference scale.

Over the years many cluster algorithms have been proposed~\cite{algrev},
each with its strengths and weaknesses. In this article we adopt the
Durham one~\cite{Durham}, which is a standard for many perturbative
calculations. The distance measure between two clusters $i$ and $j$ is
\begin{equation}
y_{ij} = \frac{2 \min(E_i^2,E_j^2)(1-\cos\theta_{ij})}{E_{\mathrm{vis}}^2}
~,
\end{equation}
so that $\sqrt{y_{ij}}$ roughly corresponds to the relative transverse
momentum, scaled to the total visible energy $E_{\mathrm{vis}}$
($= \Ecm$ for an ideal detector).

If we begin by considering a simple $q\overline{q}$ event, it should
reconstruct back to two clusters, unless $\ymin$ has been chosen very
small. Since the momentum of a cluster is given by the vector sum of
its constituent hadrons, it would resum opposite and compensating
$\pT$ kicks imparted to hadrons in the fragmentation process. The
cluster direction should therefore be a better measure of the
$q\overline{q}$ axis than that provided by the individual hadron
momenta. It is the angular distribution of this axis that relates back
to the polarization character of the $\gamma^*/Z^0 \to q\overline{q}$
decay, and that we want to be reflected in our extracted $\FL(x)$ and $\FT(x)$.
Therefore it would be an improvement to rotate all hadrons in a cluster
to sit along the cluster direction. That is, the $\theta$ of a hadron
is redefined while its $x$ value is unchanged.

At this level there is no contradiction with the smearing approach studied
earlier. Then we smeared the simple partonic angular shapes to arrive at
realistic hadronic ones to compare with data, now we un-smear the hadronic
angles to approach the simple partonic distributions. There is one
advantage, however: the clustering approach is not sensitive to the
width of the $\pT$ distribution, i.e. the $\sigma$ parameter, unlike the
smearing procedure. Of course, the $\pT$ width still affects the typical
error between the $q\overline{q}$ and cluster axes.

When considering multijet production, the $\ymin$ choice does become
relevant, with $\mu^2 = \ymin \Ecm^2$ acting as a factorization scale,
as noted above. For a large $\ymin$ all activity is clustered into the
two quark jets, and neither the gluon structure nor $\FL(x)$ would be
probed. For $\ymin \to 0$ each hadron or parton is a cluster unto
itself, and we are back at the starting point. So obviously some
intermediate scale is to be preferred. Given that the typical
hadronization $\pT$ width is $\sim 0.4$~GeV, with a tail to larger
values, one would conclude that clustering up to $\pT \sim 1$~GeV
would be a sensible minimum to eliminate the bulk of the hadronization
$\pT$ smearing. At the $Z^0$ peak this translates into $\ymin \gtrsim
0.0001$. In the upper end, we want to stay with a picture of multiple
gluon emission as the norm, i.e. retain $\FL(x)$ as an inclusive
quantity, in order not to overlap with traditional studies of gluon
jets in exclusive three-jet events. Since the average number of
clusters per event is three for $\ymin \approx 0.0025$, we conclude
that $0.0001 \lessim \ymin \lessim 0.0025$ is a reasonable range, over
which to study a scale-dependent $\FL(x,\mu)$.
\begin{figure}[ht]
\vspace*{-1cm}
\centering
\mbox{\epsfig{file=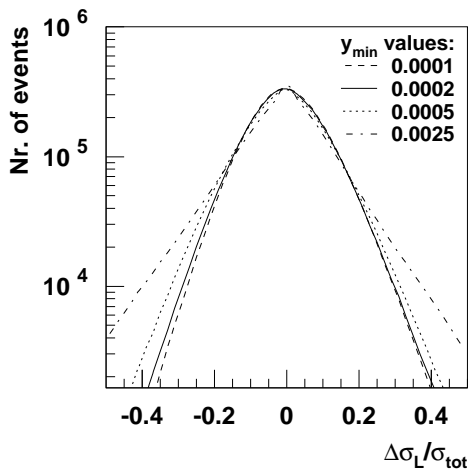,width=0.5\linewidth}} \caption{\it
  Event-by-event hadronization corrections $\Delta \sigL/\sigtot =
  (\sigL^h - \sigL^p)/\sigtot$ for different $\ymin$ scales
  (\textsc{Jetset}~7.4~PS and string fragmentation).}
\label{fig:ymindep}
\end{figure}

In Fig.~\ref{fig:ymindep}, the distribution of event-by-event
$x_E$-weigh\-ted and -integrated hadronization corrections $\Delta
\sigL/\sigtot = (\sigL^h - \sigL^p)/\sigtot$ is shown for some
different $\ymin$ scales, for events generated with parton showers and
string fragmentation (superscripts $h$ and $p$ stand for
hadron and parton level, respectively).  We note the significant width
of these distributions, showing that event-by-event fluctuations in
the hadronization process are important and can be of either sign.
Even if small by comparison, the mean $\langle \Delta \sigL/\sigtot
\rangle$ does show a systematic bias, positive for small $\ymin$ and
negative for large $\ymin$. That is, at small $\ymin$ the
hadronization smearing wins over the string effects, while it is the
other way around for large $\ymin$ --- but remember that this is only
true when averaging over many events. Nevertheless, one possible
criterion for a good choice of $\ymin$ would be where the two effects
cancel, which then gives $\ymin \approx 0.0002$, i.e. $\mu \approx
1.3$~GeV. While a sensible reference value, one should not take this
particular value too seriously, since it is for one specific model,
and for one specific set of model parameters. Somewhat different
parameter values, like for the parton shower cut-off $Q_0 \approx
1$~GeV, defining the parton level of the events studied, could lead to
slightly different ``preferred'' $\mu$ values.
\begin{figure}[hb]
\vspace*{-1cm}
\centering
\mbox{\epsfig{file=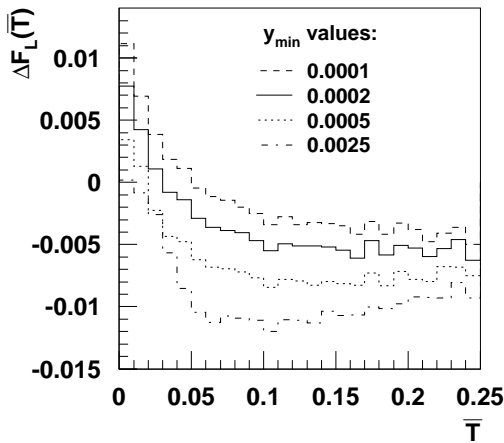,width=0.5\linewidth}} \caption{\it
  Hadronization corrections $\Delta \FL(\Tbar) = \FLh(\Tbar) - \FLp(\Tbar)$ for different $\ymin$ scales (\textsc{Jetset}~7.4~PS and string fragmentation).}
\label{fig:deltaclus}
\end{figure}

\begin{figure}[ht]
\vspace*{-1cm}
\centering
\mbox{\epsfig{file=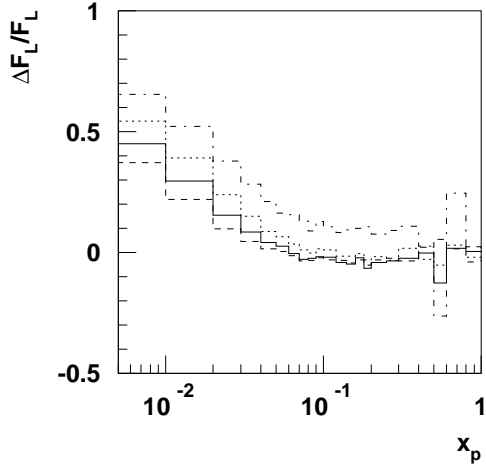,width=0.5\linewidth}} \caption{\it
 Ratio $\Delta \FL/\FL = (\FL - \FL^{cluster})/\FL$ for various $y_{min}$
 values: dashed line corresponds to
 $y_{min}=0.0001$, solid -- $y_{min}=0.0002$, dotted --
 $y_{min}=0.0005$ and dash-dotted -- $y_{min}=0.0025$.}
\label{fig:clustering}
\end{figure}

It is important to note that we here have been considering the
$x$-integrated quantity. This is of relevance if one e.g. would like
to extract an $\alphas$ from an $\sigL$ measurement, and so not
uninteresting.  For the purpose of determining the differential $x$
distribution, $\FL(x)$, however, one would have reason to fear that
any bias could have an $x$ dependence that would not be caught. In the
string model, a string piece connecting two partons is boosted by an
increasing velocity vector as the relative opening angle between the
partons is decreased, and so the string effects spread upwards to
larger $x$ values. A warning signal is then that $\Delta \FL$ does
depend quite significantly on $\Tbar$, Fig.~\ref{fig:deltaclus}, i.e.
clustering does not reduce the $\Tbar$ dependence noted in
Fig.~\ref{fig:toysim}, but mainly shifts the overall level.  Since
$\Tbar$ probes the topology of events, we also do expect this topology
to reflect itself in an $x$ dependence of hadronization corrections.
As in the previous studies, this dependence is then likely to show up
mainly in the lower end of the $x$ range. At larger $x$, hadrons are
rather well aligned with the jet axes, so, even with $x$-weigh\-ting,
the few particles out there give a small contribution to the $\langle
\Delta \FL \rangle$. In Fig~\ref{fig:clustering}, the relative difference
between the inclusive $F_L(x_p)$ and $F_L^{cluster}(x_p)$, obtained
by replacing hadron angles with cluster angles for different $y_{min}$
values, is shown. It is clearly seen, indeed, that the hadronization
corrections are only important at low $x_p$, unless $\ymin$ is chosen
too high. For the transverse
component of the fragmentation function, $F_T(x_p)$, corrections have
the same absolute amplitude but the opposite sign.

On the up side, the string effect has its perturbative
equivalent in the colour dipole~\cite{drageff}. That is, asymmetries
also exist in the production of soft gluons around the direction
of the harder partons of an event. Such soft parton emission, below
the cut-off scale $Q_0 \approx 1$~GeV we have used, would thus largely
fill in the same regions as the nonperturbative hadron production,
and with the same topology dependence. If one takes Local
Parton-Hadron Duality~\cite{LPHD} seriously, this equivalence should
come very close. Our proposed strategy, to reset the $\theta$ angle of
particles to that of the cluster they belong to, would be applicable
also to such perturbatively calculated parton topologies.

\section{Summary}

The coherence phenomenon~\cite{coherence} kills the concept of gluon
fragmentation functions that can be defined independently of the
environment they are found in. The ``hump-backed'' shape of inclusive
$x$ spectra~\cite{LPHD,hump} is an excellent illustration: by coherence
the multiplication of partons/hadrons at small $x$ is much less than
if the hard partons could radiate/hadronize independently. The
immediate consequence is that the expected ``softer gluon than quark
jets'' picture is difficult to test. This impacts both on studies of
gluon jets directly in identified three-jet events and indirectly via
$\FL(x)$. In this article we have illustrated some of these issues
for the latter observable.

It appears safe to conclude that a straightforward extraction of
$\FL(x)$ from hadron angular distributions exaggerates the rate of
particles at small $x$ that should be attributed to gluon jets,
since even the hadronization of pure $q\overline{q}$ events induces
a `false' $\FL(x)$ by $\pT$ smearing. We have also here shown that
a symmetric smearing around jet axes introduces a bias in the other
direction, since it misses important string/drag effects that tend
to make three-jet events more two-jetlike. In summary, there is no
model-independent extraction of a unique $\FL(x)$, especially not at
small $x$ values.

We therefore propose to introduce a scale-dependent quantity
$\FL(x, \mu^2)$. Particles are clustered, e.g. with the Durham
algorithm, and thereafter assigned the $\theta$ angle of the cluster
they belong to, while retaining their $x$ value. Thus $\mu^2$
sets an `un-smearing' scale, below which $\pT$ fluctuations are killed.
We find that a $\mu \approx 1.3$~GeV gives opposite and compensating
$\pT$ smearing and string effects in \textsc{Jetset} simulations.
While the exact number certainly is model-dependent, the order is a
sensible one, given that the average hadronization $\pT$ is of the
order of 0.4~GeV. If this then sets a reasonably lower limit, an upper
one is related to the desire to stay away from the region of exclusive
two- or three-jet events. Over an intermediate range, one could imagine
several experimental determinations providing the scale dependence. We
also remind that, so far, our studies have only been intended for LEP1
energies. Coverage of a wider energy range, e.g. at LEP2, introduces
$s$ as a further scale of the process and allows more differential
tests.

\end{document}